\documentclass[twocolumn]{aastex631}
\usepackage{amsmath}

\usepackage[T1]{fontenc} 
\usepackage[utf8]{inputenc} 
\usepackage{amssymb}
\usepackage{soul}

\newcommand{\ba}{\begin{eqnarray}}
\newcommand{\ea}{\end{eqnarray}}

\shorttitle{}
\shortauthors{}

\begin{document}

\title{HATS-38 b and WASP-139 b join a growing group of hot Neptunes on polar orbits\footnote{Based on observations made with ESO Telescopes at the La Silla Paranal Observatory under programs ID 111.24VT.001, 111.24VT.002, and 112.25W1.001}}

\correspondingauthor{Juan I. Espinoza-Retamal}
\email{jiespinozar@uc.cl}

\author[0000-0001-9480-8526]{Juan I. Espinoza-Retamal}
\affiliation{Instituto de Astrof\'isica, Pontificia Universidad Cat\'olica de Chile, Av. Vicu\~na Mackenna 4860, 782-0436 Macul, Santiago, Chile}
\affiliation{Millennium Institute for Astrophysics, Santiago, Chile}
\affil{Anton Pannekoek Institute for Astronomy, University of Amsterdam, Science Park 904, 1098 XH Amsterdam, The Netherlands}

\author[0000-0001-7409-5688]{Guðmundur Stefánsson} 
\affil{Anton Pannekoek Institute for Astronomy, University of Amsterdam, Science Park 904, 1098 XH Amsterdam, The Netherlands}

\author[0000-0003-0412-9314]{Cristobal Petrovich}
\affiliation{Instituto de Astrof\'isica, Pontificia Universidad Cat\'olica de Chile, Av. Vicu\~na Mackenna 4860, 782-0436 Macul, Santiago, Chile}
\affiliation{Millennium Institute for Astrophysics, Santiago, Chile}
\affiliation{Department of Astronomy, Indiana University, Bloomington, IN 47405, USA}

\author[0000-0002-9158-7315]{Rafael Brahm}
\affiliation{Millennium Institute for Astrophysics, Santiago, Chile}
\affiliation{Facultad de Ingeniería y Ciencias, Universidad Adolfo Ibáñez, Av. Diagonal las Torres 2640, Peñalolén, Santiago, Chile}
\affiliation{Data Observatory Foundation, Santiago, Chile}

\author[0000-0002-5389-3944]{Andr\'es Jord\'an}
\affiliation{Millennium Institute for Astrophysics, Santiago, Chile}
\affiliation{Facultad de Ingeniería y Ciencias, Universidad Adolfo Ibáñez, Av. Diagonal las Torres 2640, Peñalolén, Santiago, Chile}
\affiliation{Data Observatory Foundation, Santiago, Chile}
\affil{El Sauce Observatory --- Obstech, Coquimbo, Chile}

\author[0000-0002-7444-5315]{Elyar Sedaghati} 
\affiliation{European Southern Observatory (ESO), Av. Alonso de Córdova 3107, 763 0355 Vitacura, Santiago, Chile}

\author[0000-0002-0080-4565]{Jennifer \ P. Lucero} 
\affiliation{Instituto de Astrofísica e Ciências do Espaço, Universidade do Porto, CAUP, Rua das Estrelas, 4150-762 Porto, Portugal}
\affiliation{Departamento de Física e Astronomia, Faculdade de Ciências, Universidade do Porto, Rua do Campo Alegre, 4169-007 Porto, Portugal}

\author[0009-0004-8891-4057]{Marcelo Tala Pinto} 
\affiliation{Millennium Institute for Astrophysics, Santiago, Chile}
\affiliation{Facultad de Ingeniería y Ciencias, Universidad Adolfo Ibáñez, Av. Diagonal las Torres 2640, Peñalolén, Santiago, Chile}

\author[0000-0003-2186-234X]{Diego J. Muñoz} 
\affiliation{Department of Astronomy and Planetary Science, Northern Arizona University, Flagstaff, AZ 86011, USA}

\author[0009-0009-2966-7507]{Gavin Boyle} 
\affil{El Sauce Observatory --- Obstech, Coquimbo, Chile}
\affil{Cavendish Laboratory, J J Thomson Avenue, Cambridge, CB3 0HE, UK}

\author[0000-0002-6477-1360]{Rodrigo Leiva} 
\affiliation{Millennium Institute for Astrophysics, Santiago, Chile}
\affiliation{Instituto de astrofísica de Andalucía, CSIC, Glorieta de la Astronomía s/n, 18008 Granada, Spain}

\author[0000-0001-7070-3842]{Vincent Suc} 
\affiliation{Millennium Institute for Astrophysics, Santiago, Chile}
\affiliation{Facultad de Ingeniería y Ciencias, Universidad Adolfo Ibáñez, Av. Diagonal las Torres 2640, Peñalolén, Santiago, Chile}
\affil{El Sauce Observatory --- Obstech, Coquimbo, Chile}

\begin{abstract}
We constrain the sky-projected obliquities of two low-density hot Neptune planets, HATS-38 b and WASP-139 b, orbiting nearby G and K stars using Rossiter-McLaughlin (RM) observations with VLT/ESPRESSO, yielding $\lambda = -108_{-16}^{+11}$ deg and $-85.6_{-4.2}^{+7.7}$ deg, respectively. To model the RM effect, we use a new publicly available code, \texttt{ironman}, which is capable of jointly fitting transit photometry, Keplerian radial velocities, and RM effects. WASP-139 b has a residual eccentricity $e=0.103_{-0.041}^{+0.050}$ while HATS-38 b has an eccentricity of $e=0.112_{-0.070}^{+0.072}$, which is compatible with a circular orbit given our data. Using the obliquity constraints, we show that they join a growing group of hot and low-density Neptunes on polar orbits. We use long-term radial velocities to rule out companions with masses $\sim 0.3-50$ $M_J$ within $\sim10$ au. We show that the orbital architectures of the two Neptunes can be explained with high-eccentricity migration from $\gtrsim 2$ au driven by an unseen distant companion. If HATS-38b has no residual eccentricity, its polar and circular orbit can also be consistent with a primordial misalignment. Finally, we performed a hierarchical Bayesian modeling of the true obliquity distribution of Neptunes and found suggestive evidence for a higher preponderance of polar orbits of hot Neptunes compared to Jupiters. However, we note that the exact distribution is sensitive to the choice of priors, highlighting the need for additional obliquity measurements of Neptunes to robustly compare the hot Neptune obliquity distribution to Jupiters.
\end{abstract}

\keywords{Exoplanets (498) --- Hot Neptunes (754) --- Exoplanet dynamics (490) --- Planetary alignment (1243)}

\section{Introduction}
\label{sec:intro}
More than 5,000 exoplanets have been confirmed, with super-Earths and mini-Neptunes being the most prevalent. Essential information about the formation of these planetary systems can be found in their architectures, as they serve as a powerful probe of their dynamical history. One of the main signatures of the dynamical evolution of systems is the stellar obliquity ($\psi$)---the angle between the stellar spin axis and the planet's orbital axis. Significant progress has been made in recent years in measuring the sky-projected obliquity ($\lambda$) distribution of hot Jupiters. Measurements performed via the Rossiter-McLaughlin (RM) effect have uncovered a broad distribution of $\lambda$ for hot Jupiters, ranging from well-aligned systems to systems on highly misaligned orbits \citep{Albrecht2012}. These results have been interpreted as evidence that hot Jupiter formation involves dynamical perturbations that excite orbital inclinations. 

However, the signature of the dynamical evolution is the true 3D obliquity $\psi$, not its sky projection $\lambda$.
The true obliquity $\psi$ can be estimated from $\lambda$ measured by the RM effect through combining it with constraints on the orbital inclination, $i$ constrained by the shape of a planetary transit, and the stellar inclination, $i_{\star}$, which can be constrained by measuring the stellar rotation period ($P_{\rm rot}$), the projected rotational velocity ($v\sin i_\star$), and the radius of the host star \citep[e.g.,][]{Hirano2014,Morton2014,Masuda2020}. 

By systematically calculating true obliquities from sky-projected obliquities, \citet{Albrecht2021} revealed an architectural dichotomy: hot Jupiter orbits appear to separate into well-aligned ($\psi\sim0$ deg) and polar ($\psi\sim90$ deg) orbits. However, more recent work by both \citet{Siegel2023} and \citet{Dong23}, using Bayesian approaches, did not find strong evidence for a dichotomy. In turn, the Bayesian modeling suggests the $\psi$ distribution is unimodal and peaked at $0$ deg with an almost isotropic tail for misaligned systems, with no significant clustering at $90$ deg. As hot Jupiters represent an infrequent outcome of planet formation \citep{Batalha2013}, it is unclear if the observed obliquity distribution is related to the formation of hot Jupiters, or instead reflects more general aspects of star and planet formation.

Thanks to recent observational efforts, the current sample of planets with measured obliquity shows a growing number of smaller planets including hot Neptunes ($10<M_p/M_{\oplus}<50$ or $2<R_p/R_{\oplus}<6$) in the ``hot Neptune desert'', a region of the period-radius diagram where Neptune-sized exoplanets are notably rare \citep{Mazeh2016} likely due to intense stellar radiation causing atmospheric erosion. Interestingly, many hot Neptunes are observed to be in close-to-polar orbits and have evaporating atmospheres, including HAT-P-11 b \citep{Sanchis-Ojeda2011}, GJ 436 b \citep{Bourrier2018,Bourrier2022}, WASP-107 b \citep{Dai2017,Rubenzahl2021}, and GJ 3470 b \citep{Stefansson22}, suggesting a possible link between the Neptune desert and the dynamical history of low-mass gas giants. Theoretical scenarios that are capable of explaining the polar orbits of these planets require the presence of massive distant companions \citep[e.g.,][]{Petrovich2020}. Recently, outer companions have been confirmed in the HAT-P-11 \citep{Yee2018} and WASP-107 \citep{Piaulet2021} systems, suggesting that the transiting planets arrived at their current polar orbits through dynamical interactions \citep[see][]{Lu_ZLK_2024,YU2024}. As for the atmospheres, the orbital eccentricities may be large enough so that tidal heating can inflate these planets \citep[e.g., by a factor of $\simeq1.5-1.7$ for GJ 3470 b;][]{Stefansson22}, possibly accounting for their evaporation. Nevertheless, the sample of hot Neptunes with measured obliquities is still small, and more measurements are necessary to test these emerging trends and asses the underlying mechanism behind the formation of this transiting population.

In this work, we present ESPRESSO \citep{Pepe20} observations of the RM effect of the two hot Neptunes HATS-38 b \citep{Jordan2020} and WASP-139 b \citep{Hellier2017}. HATS-38 b is a hot Neptune in the middle of the hot Neptune desert orbiting a G-dwarf host star, while WASP-139 b is a super-Neptune orbiting a K-star. The precise in-transit spectroscopic observations with ESPRESSO reveal that both planets are in nearly polar orbits, continuing with the observed trends for low-density gas giants. We also performed a long-term radial velocity (RV) monitoring of both stars in order to constrain the presence of distant planetary companions in the systems.

We describe our observations in Section \ref{sec:obs}. In Section \ref{sec:star}, we present our stellar analysis for both stars. In Section \ref{sec:photo}, we update the orbital ephemeris of HATS-38 b and WASP-139 b, while in Section \ref{sec:fit}, we describe the joint analysis of photometry and RV data to get the parameters of the planets and their orbits. We discuss the implications of these measurements in Section \ref{sec:discussion} and summarize our findings in Section \ref{sec:conclusion}.

\begin{figure*}
\begin{center}
    \includegraphics[scale=0.8]{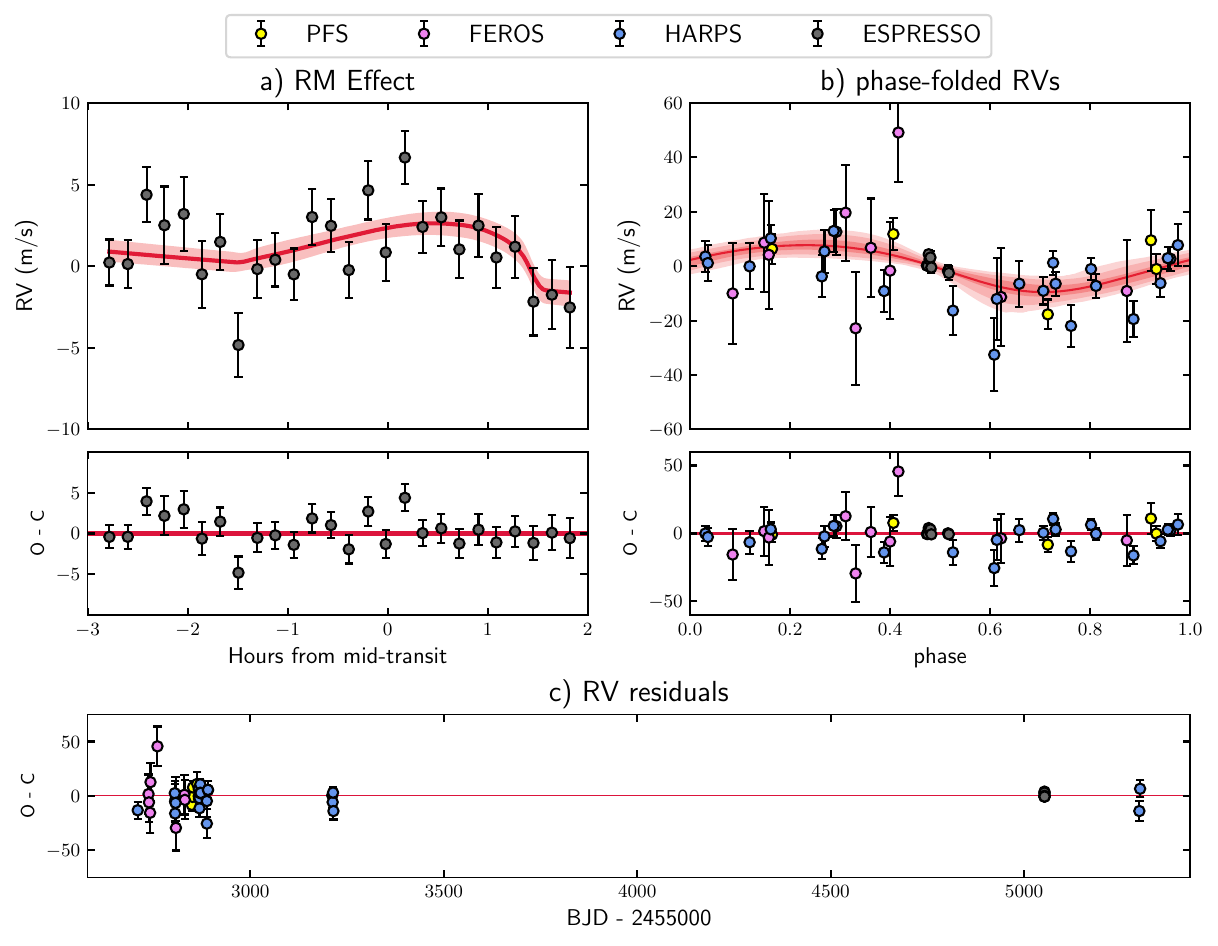}
    \includegraphics[scale=0.8]{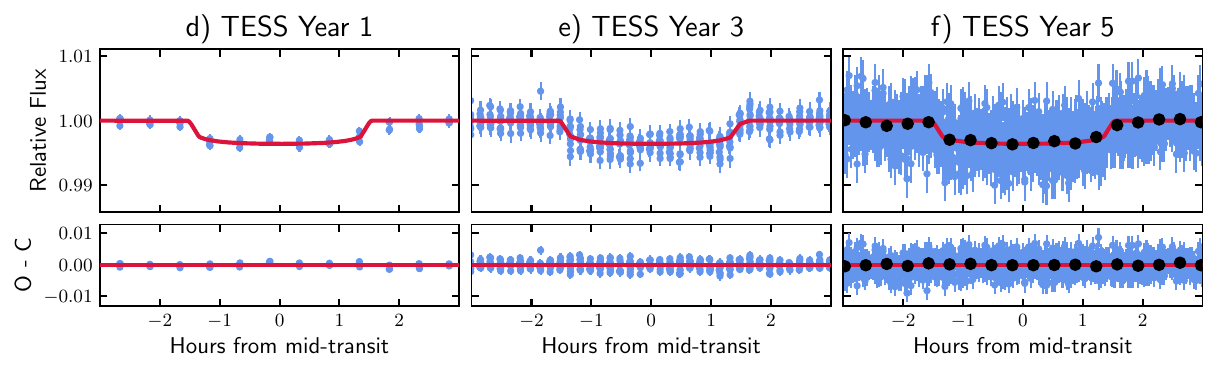}
    \includegraphics[width=\textwidth]{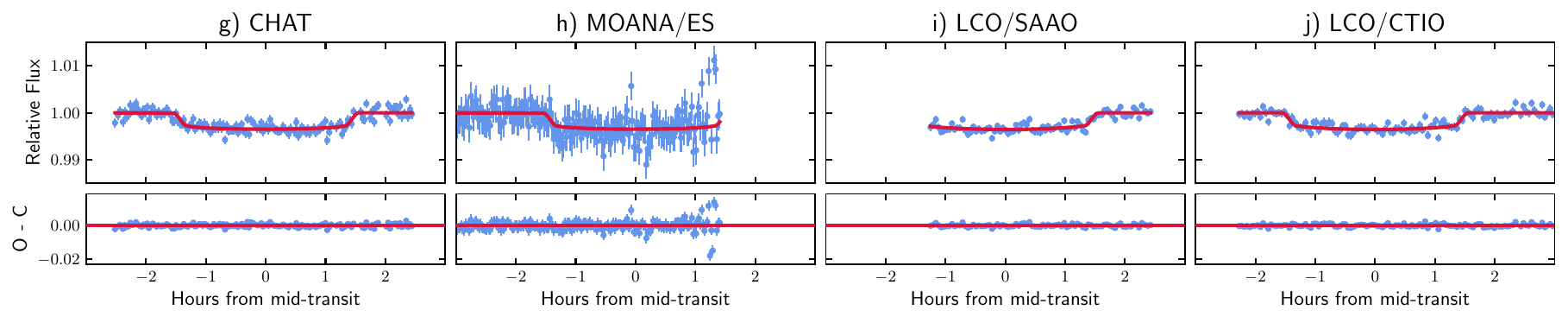}
    \caption{Observations of HATS-38 along with the best-fit model. All error bars include a white noise jitter term. a) ESPRESSO observations of the RM effect. The red line and shaded area correspond to the best and $1\sigma$ models, respectively. b) Phase-folded out-of-transit RVs showing $1\sigma$, $2\sigma$, and $3\sigma$ models. c) Residuals of the model for the RVs as a function of time. d-j) Different light curves. We show the binned TESS Year 5 data in black.}
    \label{fig:HATS38}
\end{center}
\end{figure*}

\begin{figure*}
    \centering
    \includegraphics[scale=0.8]{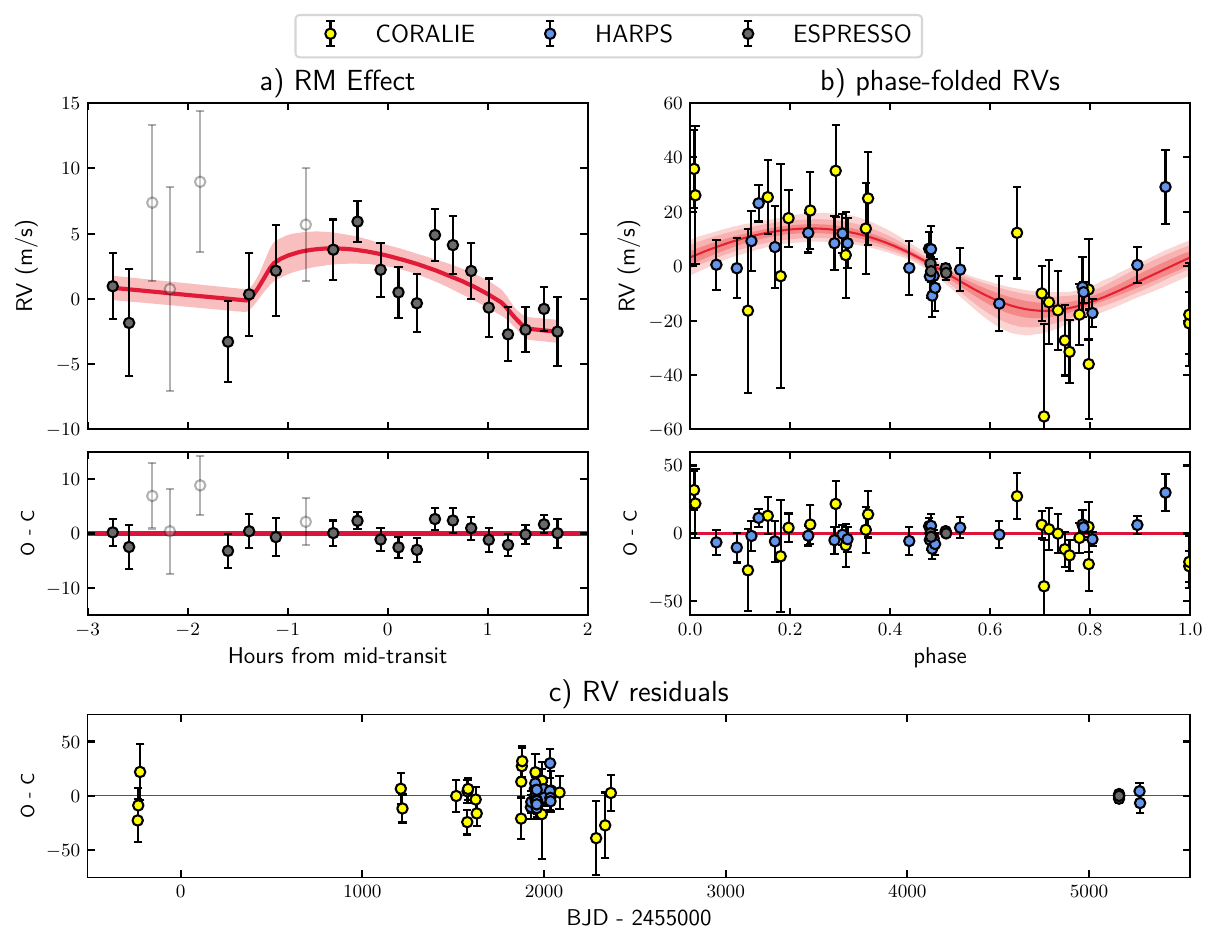}
    \includegraphics[scale=0.8]{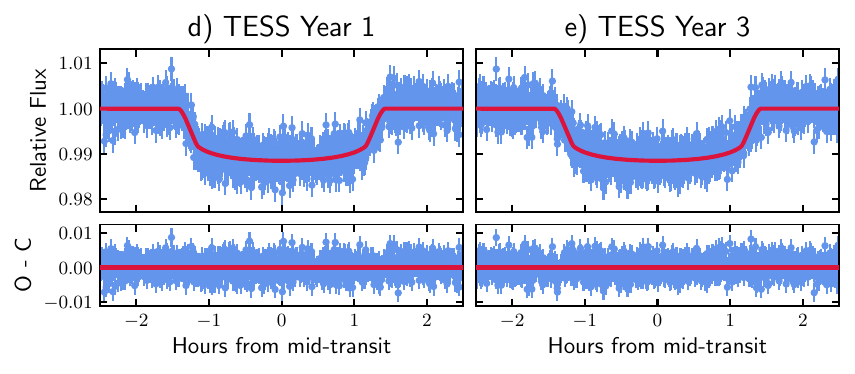}
    \includegraphics[scale=0.8]{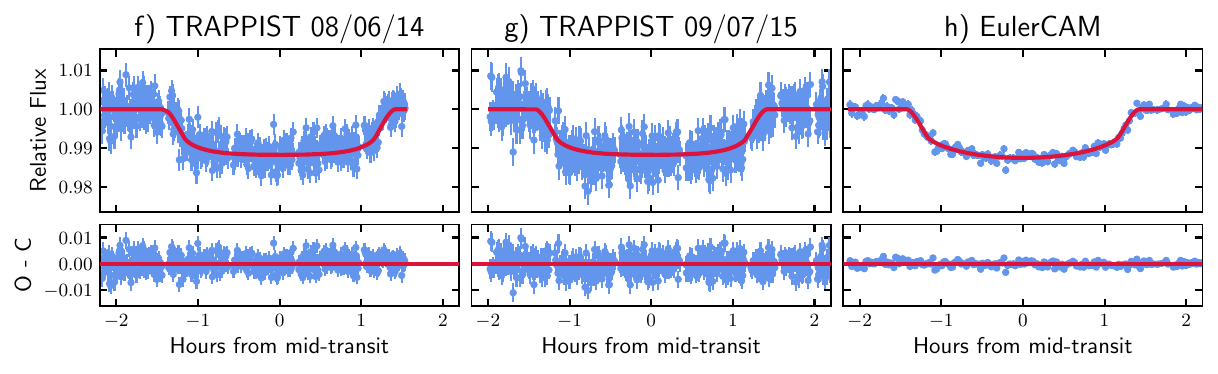}
    \caption{Same as Figure \ref{fig:HATS38} but for WASP-139. White points in the RM effect plot were not considered for the fit given their low S/N due to increased cloud coverage during those exposures. Including or removing those points results in the same constraints on the sky-projected obliquity.}
    \label{fig:WASP139}
\end{figure*}

\begin{deluxetable*}{lccr}
\tablecaption{Summary of observations used in this work.  \label{tab:obs}}
\tablecolumns{4}
\tablewidth{0pt}
\tablehead{Facility & Date & Notes & Reference}
\startdata
\textbf{HATS-38} &  &  &\\
ESPRESSO & 2023 Apr 18 & 26 RVs & This work\\
HARPS$^a$ & 2016 Nov–2017 May & 24 RVs& \citet{Jordan2020}\\
FEROS & 2016 Dec–2017 Mar & 10 RVs  & \citet{Jordan2020}\\
PFS & 2017 Apr 05-19 & 5 RVs & \citet{Jordan2020}\\
TESS$^b$ & Years 1, 3 \& 5  & 30, 10 \& 2 min cadence & This work\\
MOANA/ES & 2023 Apr 18  & $r'$ filter & This work \\
CHAT & 2017 Feb 05  & $r'$ filter & \citet{Jordan2020} \\
LCO/SAAO  &2017 Mar 30	& $i'$ filter & \citet{Jordan2020} \\
LCO/CTIO  &2017 Apr 03	& $i'$ filter & \citet{Jordan2020} \\
\hline
\textbf{WASP-139} &  &  &\\
ESPRESSO & 2023 August 08 & 22 RVs & This work\\
HARPS$^a$ & 2014 Sep-2015 Jan & 23 RVs& \citet{Hellier2017}\\
CORALIE$^c$ & 2008 Oct–2015 Dec & 24 RVs & \citet{Hellier2017}\\
TESS$^b$ & Years 1 \& 3  & 2 min cadence & This work\\
TRAPPIST & 2014 Aug 06  & $I + z$ band & \citet{Hellier2017} \\
TRAPPIST & 2015 Sep 07  & $I + z$ band & \citet{Hellier2017} \\
EulerCAM & 2015 Sep 07  & $NGTS$ filter & \citet{Hellier2017}\\
\enddata
\tablecomments{$^a$ We took two extra HARPS measurements per target and reduced all data again in a uniform way (see Section \ref{sec:harps}).$^b$ We combined all the TESS sectors of each year, and in the analysis, we assumed that different years were different instruments.$^c$ We considered the pre and post CORALIE update data as taken by different instruments.}
\end{deluxetable*}

\section{Observations}\label{sec:obs}

Table \ref{tab:obs} shows a summary of the observations used in this work. Next, we describe novel observations of HATS-38 and WASP-139. 

\subsection{ESPRESSO Transit Spectroscopy}

We observed a single transit of each planet, HATS-38 b and WASP-139 b, with the ESPRESSO spectrograph \citep{Pepe20}. ESPRESSO is a highly-stabilized, fiber-fed cross-dispersed echelle spectrograph installed at the Incoherent Combined Coudé Focus of ESO's Paranal Observatory in Chile. It covers a wavelength range from 380 to 788 nm at a resolving power of $R\approx$ 140,000 in single Unit Telescope (UT) high-resolution mode. The transit of HATS-38 b was observed on the night of 2023 April 18, between 01:25 and 06:01 UTC. We obtained 26 spectra of the host star during the primary transit with UT1 at an exposure time of 610 s. The observations were performed under clear sky conditions, with atmospheric seeing in the range of $0.42-1.7^{\prime\prime}$. The spectra have a median signal-to-noise ratio (S/N) of 39 at 550 nm and a median RV uncertainty of 1.7 m/s. The transit of WASP-139 b was observed on the night of 2023 August 8, between 05:35 and 10:04 UTC. We obtained 22 spectra of the host star during the primary transit with UT3. The sky was cloudy at the beginning of observations before the transit, and conditions improved during the night. We started observations with an exposure time of 900 s, and then changed it to 600 s when conditions improved. In our analysis, we excluded 4 of the 22 spectra given their low S/N ($<20$ at 550 nm). The remaining 18 spectra have a median S/N of 31  at 550 nm and a median RV uncertainty of 2.2 m/s. Both datasets were reduced with the ESO Reflex environment \citep{Freudling13} using the dedicated ESPRESSO data reduction pipeline \citep[v. 3.1.0,][]{Sosnowska15,Modigliani20}, including all standard reduction steps, which also provides RVs by fitting a Gaussian model to the calculated cross-correlation function (CCF). The CCF was calculated at steps of 0.5 km/s (representing the sampling of the spectrograph) for $\pm 20$ km/s centered on the estimated systemic velocity. The RVs showing the RM effect of both targets are presented in panels a of Figures \ref{fig:HATS38} and \ref{fig:WASP139}.

\subsection{HARPS Follow-up Spectroscopy}\label{sec:harps}

HATS-38 and WASP-139 were both observed with the HARPS spectrograph mounted on the 3.6m telescope installed at the ESO La Silla Observatory. HARPS is a high-resolution ($R\approx$ 120,000) stabilized instrument that uses a secondary fiber to trace instrumental velocity variations by measuring drifts of a Fabry-Perot interferometry pattern. Archival observations were available through the ESO archive. These observations were used initially to confirm the planetary nature of both planets. Besides these archival HARPS data, we obtained two additional spectra for each system in order to look for long-period companions. These new observations were performed in November and December of 2023. For HATS-38 we used an exposure time of 1500 s while for WASP-139 we adopted an exposure time of 1800 s. Archival and new HARPS data were homogeneously processed with the \texttt{ceres} \citep{ceres} pipeline, which performs all processing steps to obtain precision RVs starting from the raw images. 

\subsection{Observatoire Moana Photometry}  

Simultaneously with ESPRESSO observations, we observed the transit of HATS-38 b using the station of the Observatoire Moana located in El Sauce (ES) Observatory in Chile. Observatoire Moana is a global network of small-aperture robotic optical telescopes. The ES station consists of a 0.6 m CDK robotic telescope coupled to an Andor iKon-L 936 deep depletion 2k $\times$ 2k CCD with a scale of 0.67$^{\prime\prime}$ per pixel. For these observations, we used a Sloan $r'$ filter, and the exposure time was set to 74 s. Data reduction was done using a dedicated pipeline that automatically performs the CCD reduction steps, followed by the measurement of the aperture photometry for the brightest stars in the field. The pipeline also generates the differential light curve of the target star by identifying the optimal comparison stars based on color, brightness, and proximity to the target. The MOANA/ES light curve is shown in panel h of Figure \ref{fig:HATS38}, along with the best model.

\subsection{TESS Photometry}

For this work, we also used all the available TESS \citep{Ricker2015} light curves of both targets. HATS-38 was observed in Sector 9 (Year 1) at a cadence of 30 minutes, in Sectors 35 and 36 (Year 3) at a cadence of 10 minutes, and in Sector 62 (Year 5) at a cadence of 2 minutes. WASP-139 was observed by TESS in Sectors 3 and 4 (Year 1), and 31 and 32 (Year 3) at a cadence of 2 minutes. We searched and downloaded all the light curves using \texttt{lightkurve} \citep{Lightkurve}. For this work, we used TESS light curves that were processed by the Science Processing Operations Center (SPOC) pipeline \citep{Jenkins16}. In particular, we used the Presearch Data Conditioning (PDC) light curves of the SPOC pipeline, which are corrected for pointing or focus-related instrumental signatures, discontinuities resulting from radiation events in the CCD detectors, outliers, and flux contamination. TESS PDC light curves, together with the best model, are shown in the bottom panels of Figures \ref{fig:HATS38} and \ref{fig:WASP139}.

\section{Stellar Parameters}\label{sec:star}

We obtained the stellar parameters of HATS-38 and WASP-139 following the procedure presented in \citet{Brahm2019}. In brief, we used a two-step iterative process. 

The first process consists of obtaining the atmospheric parameters ($T_{\rm eff}$, $\log{g}$, [Fe/H], and $v\sin{i}$) of the host star from a high-resolution spectrum using the \texttt{zaspe} code \citep{zaspe}. This code compares the observed spectrum to a grid of synthetic spectra and determines reliable uncertainties that take into account systematic mismatches between the observations and the imperfect theoretical models. For this analysis, we used the co-added out-of-transit HARPS spectra to obtain the atmospheric parameters. 

The second step consists of obtaining the physical parameters of the star by using publicly available broad-band magnitudes of the star and comparing them with those produced by different PARSEC stellar evolutionary models \citep{parsec} by taking into account the distance to the star computed from the Gaia DR2 \citep{dr2} parallax. This procedure delivers a new value of $\log{g}$ that is held fixed in a new run of \texttt{zaspe}. We iterate between the two procedures until reaching convergence, which happens when two consecutive \texttt{zaspe} runs deliver the same values of $T_{\rm eff}$ and [Fe/H]\footnote{these parameters come from a grid search, so that for small changes in $\log{g}$ ($\Delta \log{g} < 0.01$) the model with the best matching $T_{\rm eff}$ and [Fe/H] will be the same.} The obtained parameters for each star are presented in Table \ref{tab:stellar}.

\begin{deluxetable}{lccr}
\tablecaption{Stellar properties$^a$ of HATS-38 and WASP-139.  \label{tab:stellar}}
\tablecolumns{4}
\tablewidth{0pt}
\tablehead{Parameter & HATS-38 & WASP-139 & Reference}
\startdata
RA \dotfill (J2015.5) & 10h17m05.05s & 03h18m14.92s & Gaia DR2\\
Dec \dotfill (J2015.5) & -25h16m34.67s & -41d18m07.28s & Gaia DR2\\
pm$^{\rm RA}$ \hfill (mas/yr) & -21.75$\pm$0.07 & -15.99$\pm$0.03 & Gaia DR2\\
pm$^{\rm DEC}$ \dotfill (mas/yr) & -7.54$\pm$0.07 & 24.51$\pm$0.05 & Gaia DR2\\
$\pi$ \dotfill (mas)& 2.88$\pm$0.04& 4.68$\pm$0.02 & Gaia DR2 \\ 
\hline
T \dotfill (mag) & 11.813$\pm$0.007 & 11.728$\pm$0.006 & TICv8\\
B  \dotfill (mag) & 12.76$\pm$0.42 & 13.330$\pm$0.016 &  APASS$^b$\\
V  \dotfill (mag) &  11.967$\pm$0.030& 12.456$\pm$0.046 & APASS\\
G  \dotfill (mag) & 12.278$\pm$0.0002 & 12.271$\pm$0.0001 &  Gaia DR2$^c$\\
G$_{BP}$  \dotfill (mag) &  12.649 $\pm$ 0.0002  & 12.733 $\pm$ 0.0002 & Gaia DR2\\
G$_{RP}$  \dotfill (mag) &  11.761 $\pm$ 0.0002  & 11.677 $\pm$ 0.0002 & Gaia DR2\\
J  \dotfill (mag) & 11.184$\pm$0.026  & 10.982$\pm$0.023 & 2MASS$^d$\\
H  \dotfill (mag) & 10.850$\pm$0.024 &  10.575$\pm$0.023 &  2MASS\\
K$_s$  \dotfill (mag) & 10.768$\pm$0.024 & 10.472$\pm$0.021 & 2MASS\\
\hline
$T_{\rm eff}$  \dotfill (K) & 5662 $\pm$ 80 &5233$\pm$60  & This work\\
$\log{g}$ \dotfill (dex) & 4.34$\pm$0.02 & 4.56 $\pm$ 0.02& This work\\
$[$Fe/H$]$ \dotfill (dex) & 0.0 $\pm$ 0.05& 0.03$\pm$0.05 & This work\\
$v\sin{i}$ \dotfill (km/s) & 3.26$\pm$0.3 & 2.49 $\pm$0.3 & This work\\
$M_{\star}$ \dotfill ($M_{\odot}$) & 0.92$\pm$0.03& 0.84$\pm$0.03 & This work\\
$R_{\star}$ \dotfill ($R_{\odot}$) &1.08$\pm$0.02 & 0.810$\pm$0.008& This work\\
L$_{\star}$ \dotfill (L$_{\odot}$) & 1.11$\pm$0.06& 0.44$\pm$ 0.02 & This work\\
A$_{V}$ \dotfill (mag) &0.12$\pm$0.08 &0.09$\pm$0.06 & This work\\
Age \dotfill (Gyr) & 10.3$_{-2.1}^{+1.6}$& 6.2$^{+3.1}_{-3.4}$ & This work\\
$\rho_\star$ \dotfill (g/cm$^{3}$) & 1.02$_{-0.05}^{+0.08}$& 2.25$\pm$0.13 & This work\\
\enddata
\tablecomments{$^a$ The stellar parameters computed in this work do not consider possible systematic differences among different stellar evolutionary models \citep{tayar:2022} and have underestimated uncertainties, $^b$\citet{apass}, $^c$\citet{dr2}, $^d$\citet{2mass}.}
\end{deluxetable}

\section{Photometric Analysis}\label{sec:photo}

In order to update the orbital ephemeris of HATS-38 b and WASP-139 b, and look for Transit Timing Variations (TTVs), we performed a photometric analysis with \texttt{juliet} \citep{Espinoza19} for each planet. \texttt{juliet} uses \texttt{batman} \citep{Kreidberg15} for the transit model and the \texttt{dynesty} dynamic nested sampler \citep{Speagle20} to perform bayesian analysis and explore the likelihood space to obtain posterior probability distributions. We placed uninformative priors on the transit parameters $R_p/R_{\star}$ and $b$, with an informative prior on the stellar density, that was constrained in Section \ref{sec:star}. We sampled the limb darkening parameters using the quadratic $q_1$ and $q_2$ limb darkening formalism from \citet{Kipping13} with uniform priors. We placed Gaussian priors for each transit mid-point based on the expected values calculated from the orbital period and $t_0$ from \citet{Jordan2020} and \citet{Hellier2017} for HATS-38 b and WASP-139 b, respectively, placing a large width of 0.1 days on the Gaussian prior standard deviation to not impact the derived transit midpoints. To account for variability and systematic noise in the TESS light curves, we included a Matern-3/2 Gaussian Process (GP) as implemented in \texttt{celerite} \citep{Foreman-Mackey17} and available in \texttt{juliet}. Each TESS year (i.e., all combined sectors of each year) had its own GP kernel to account for differences in variability captured in different epochs and cadences. For the MOANA/ES light curve, we also included a linear trend using the airmass as input as the observed flux was seen to be correlated with the airmass of the observations. From this analysis, we ruled out the presence of TTVs greater than $\sim10$ minutes for HATS-38 b and $\sim3$ minutes for WASP-139 b. Additionally, we obtained detrended TESS and MOANA/ES light curves, and ephemeris consistent with previous efforts that used different input data from HATSouth \citep{Bakos2013} and WASP \citep{Pollacco2006}.

\section{Joint fit}\label{sec:fit}

To precisely constrain the parameters of HATS-38 b and WASP-139 b, and their orbits, we broadly followed the methodology of \citet{EspinozaRetamal2023}, which we have implemented in a code named \texttt{ironman} that is publicly available on both GitHub\footnote{\url{https://github.com/jiespinozar/ironman}} and Zenodo \citep{ironman-zenodo}. Since this is the first time we use the code, we describe it here in detail. \texttt{ironman} is a Python-based code that allows the user to fit transit light curves, Keplerian RVs, and the RM effect, all together in order to better constrain the parameters of a given system. To model the transit light curves, \texttt{ironman} uses \texttt{batman} \citep{Kreidberg15}. The default parameters required to model a light curve are the orbital period ($P$), time of mid-transit ($t_0$), radius ratio ($R_p/R_{\star}$), orbital inclination ($i$), scaled semimajor axis ($a/R_{\star}$), eccentricity ($e$), argument of periastron ($\omega$), limb darkening coefficients of the instrument/band ($q_1^{\rm inst}$ and $q_2^{\rm inst}$), and the jitter term ($\sigma_{\rm inst}$). In this case $q_1^{\rm inst}$ and $q_2^{\rm inst}$ are the limb darkening parameters from \citet{Kipping13}. Additionally, \texttt{ironman} can accept the impact parameter ($b$) and the stellar density ($\rho_{\star}$) instead of $i$ and $a/R_{\star}$, respectively. To model the RVs, \texttt{ironman} uses \texttt{rmfit} \citep{Stefansson20,Stefansson22}, which uses \texttt{radvel} \citep{Fulton18} to get the Keplerian orbit, and the framework from \citet{Hirano10} to model the RM effect. The required parameters to model the Keplerian orbit of the planet are $P$, $t_0$, $e$, $\omega$, RV semiamplitude ($K$), RV offset ($\gamma_{\rm inst}$), RV linear trend ($\dot{\gamma}$), RV quadratic trend ($\ddot{\gamma}$), and $\sigma_{\rm inst}$. Finally, to model the RM effect, the sky-projected obliquity ($\lambda$), the projected rotational velocity of the star ($v\sin{i_{\star}}$), and the intrinsic linewidth $\beta_{\rm inst}$---which accounts for instrumental and macroturbulence broadening---are also required. Further, \texttt{ironman} can also accept a different parametrization of the RM effect, sampling the RM model using the stellar inclination ($\cos{i_{\star}}$), stellar rotational period ($P_{\rm rot}$), and stellar radius ($R_{\star}$) instead of $v\sin{i_{\star}}$, in order to constrain the stellar inclination and the true obliquity, following the parametrization used in \cite{Stefansson22} which broadly follows the parametrization from \citet{Masuda2020} to take into account the fact that the equatorial velocity and $v\sin{i_{\star}}$ are not independent variables. If this last parametrization is used, \texttt{ironman} estimates the true 3D obliquity $\psi$ as:
\begin{equation}
\cos{\psi}=\cos{i_{\star}}\cos{i} + \sin{i_{\star}}\sin{i} \cos{\lambda}.
\end{equation}

\begin{figure*}[t]
    \centering
    \includegraphics[width=\textwidth]{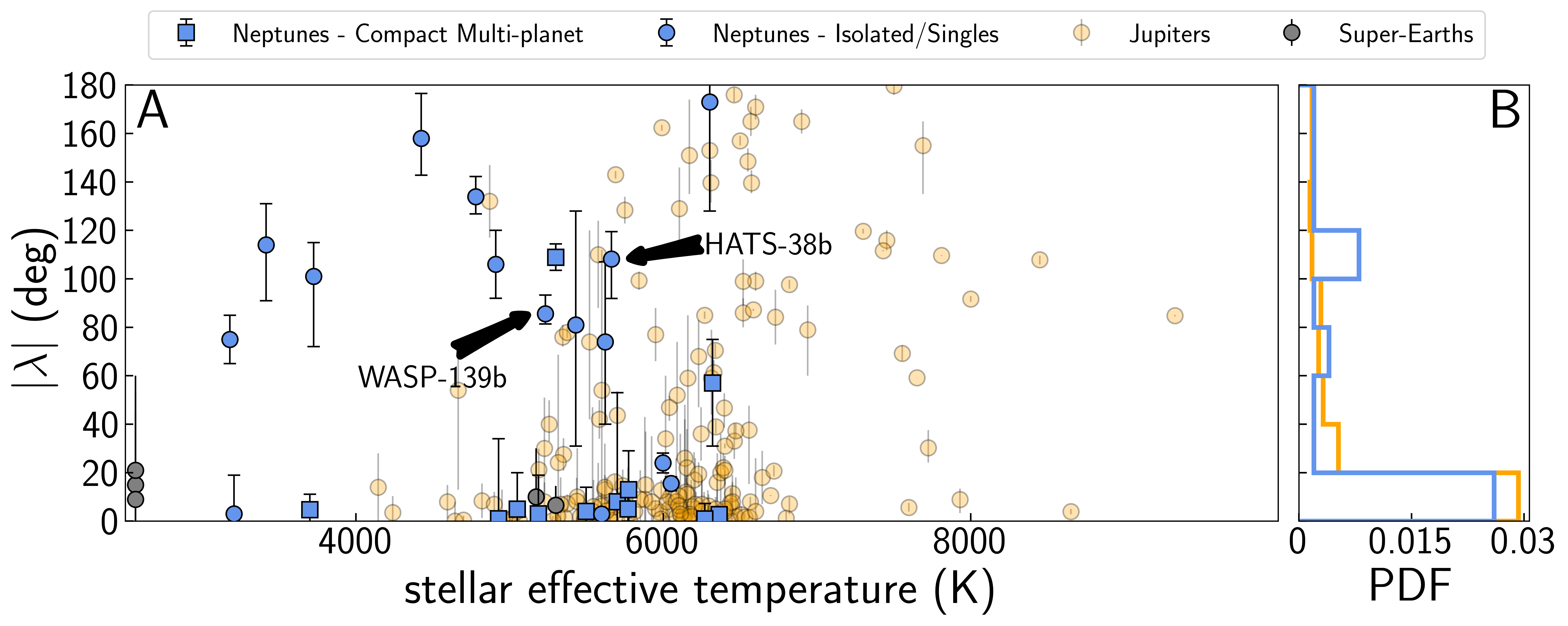}
    \caption{a) Sky-projected obliquities of Neptunes ($10<M_p/M_{\oplus}<50$ or $2<R_p/R_{\oplus}<6$) as a function of stellar effective temperature (blue). Neptunes on compact multiplanet systems are shown as squares, while isolated Neptunes are shown as circles. Jupiters (not Neptunes with $R_p/R_{\oplus}>6$) and super-Earths (not Neptunes with $R_p/R_{\oplus}<2$) are shown in orange and grey, respectively. Data from TEPCat \citep{Southworth2011}. b) Histograms of the $\lambda$ distribution for Neptunes (blue) and Jupiters (orange).}
    \label{fig:lambda_vs_teff}
\end{figure*}

To estimate the Bayesian posteriors, \texttt{ironman} uses the \texttt{dynesty} dynamic nested sampler \citep{Speagle20}. The number of live points to consider and threads to be used can be specified by the user. \texttt{ironman} accepts uniform, log-uniform, normal, and truncated normal prior distributions for the different parameters, which can also be held fixed.

With \texttt{ironman}, we analyzed both systems, including all the observations presented in Section \ref{sec:obs}. Given that we did not observe any significant variability in the TESS light curves that would allow us to measure the rotational period of the host stars, we have only fitted the sky-projected obliquity $\lambda$ and $v\sin{i_\star}$ here. For this analysis, we only considered the detrended TESS light curves (see Section \ref{sec:photo}) close to the transits, in windows of $\sim6$ hours around the transit midpoint, to reduce the computational cost. To account for the instrumentation offsets and systematics, we included independent RV offsets and log-uniform jitter terms for each instrument. We placed uninformative priors for almost all parameters, except for the orbital periods and times of mid-transit, which had been constrained in Section \ref{sec:photo}, and for the stellar densities that were constrained in Section \ref{sec:star}. Likewise for $\beta$ where we considered an instrumental broadening of 2.15 km/s because of the ESPRESSO resolution, and we used the macroturbulence law from \citet{Valenti2005}, which yields a broadening of $\sim4.4$ and $\sim3.8$ km/s for $T_{\rm eff} = 5662$  and $5233$ K, respectively. We added the instrumental and macroturbulence broadening in quadrature to set our priors for each target respectively, with an uncertainty of 2 km/s. The resulting posteriors of our fits for HATS-38 b and WASP-139 b are shown in Table \ref{tab:fit}.

Figures \ref{fig:HATS38} and \ref{fig:WASP139} show the different datasets together with the best-fit model. We found that both planets lie in a nearly-polar orbit, with sky-projected obliquities $\lambda = -108_{-16}^{+11}$ deg and $-85.6_{-4.2}^{+7.7}$ deg for HATS-38 b and WASP-139 b, respectively. Also, we found that both planets can be consistent with having eccentric orbits. This was not noted in \citet{Jordan2020} and \citet{Hellier2017}, where they imposed circular orbits. In both cases, we tested fitting circular and eccentric orbits. For WASP-139 b the Bayesian evidence ($\log{Z}$) strongly favored the eccentric model, having a Bayes factor $\Delta\log{Z} = 13.4$ and $\Delta{\rm AIC} = 21.5$, so is the fit we report in Table \ref{tab:fit}. In contrast, for HATS-38 b, we have $\Delta \log{Z} = 0.3$ and $\Delta{\rm AIC} = 2.7$, suggesting that there is not sufficient evidence to favor the eccentric model. We note that the posteriors for all the parameters are consistent between the eccentric and circular fits, where the sky-projected obliquity is $\lambda = -111^{+12}_{-13}$ deg for the circular model in agreement with the value listed in Table \ref{tab:fit}. Although the circular fit formally has 2 fewer parameters than the eccentric fit, we elected to formally adopt the values from the eccentric fit to highlight the possible range of eccentricities compatible with the data. This possible eccentricity can be further constrained with additional RV or possibly secondary eclipse observations. Additionally, we also tested sampling the eccentricity and $\omega$ parameters as $\sqrt{e}\cos{\omega}$ and $\sqrt{e}\sin{\omega}$ which resulted in a set of parameters fully consistent with the ones reported in Table \ref{tab:fit}. We also tried fitting a long-term linear RV slope to the data, but it resulted in being consistent with zero so we fixed it to that value. Overall, the rest of the parameters are consistent with the ones reported in \citet{Jordan2020} and \citet{Hellier2017}.

Finally, we evaluated the possibility of a correlation between the measured ESPRESSO RVs and stellar activity. Using \texttt{actin2} \citep{Gomes18,Gomes21} and the ESPRESSO DAS pipeline \citep[v. 1.3.7,][]{Cupani15} we calculated a number of activity indexes including H$\alpha$, $R_{HK}$, $S$, the bisector span, and the FWHM of the CCFs. We do not observe any significant correlation between the RVs and any of the indexes in both datasets.

\section{Discussion}\label{sec:discussion}

In this work, we have characterized the orbits HATS-38 b and WASP-139 b, and found that they have nearly polar sky-projected obliquities, thus joining the growing sample of isolated (without nearby companions within 4 times the orbital period) low-density Neptunes in polar orbits that include HAT-P-11 b \citep{Sanchis-Ojeda2011}, WASP-107 b \citep{Dai2017,Rubenzahl2021}, GJ 436 b \citep{Bourrier2018,Bourrier2022}, and GJ 3470 b \citep{Stefansson22}.  

\begin{deluxetable*}{llcc}
\tablecaption{Summary of posteriors of the joint fits. \label{tab:fit}}
\tablewidth{70pt}
\tabletypesize{\scriptsize}
\tablehead{Parameter & Description & HATS-38 b & WASP-139 b}
\startdata 
$\lambda$ & Sky-projected obliquity (deg) & $-108_{-16}^{+11}$ & $-85.6_{-4.2}^{+7.7}$ \\
$v\sin{i}$ & Projected rotational velocity (km/s) & $1.9_{-0.6}^{+1.1}$ & $1.8_{-0.8}^{+1.6}$ \\
$P$ & Orbital period (days) & $4.37504_{-0.00002}^{+0.00002}$ & $5.9242705_{-0.0000007}^{+0.0000008}$ \\
$t_0$ & Transit midpoint (BJD) & $2457786.4105_{-0.0005}^{+0.0005}$ & $2456870.9582_{-0.0002}^{+0.0002}$ \\
$\rho_\star$ & Stellar density (g/cm$^3$)& $1.02_{-0.08}^{+0.08}$ & $2.28_{-0.13}^{+0.13}$ \\
$b$ & Impact parameter & $0.42_{-0.21}^{+0.14}$ & $0.19_{-0.09}^{+0.11}$ \\
$R_p/R_\star$ & Radius ratio & $0.057_{-0.001}^{+0.001}$ & $0.099_{-0.001}^{+0.001}$ \\
$e$ & Eccentricity & $0.112_{-0.070}^{+0.072}$ & $0.103_{-0.041}^{+0.050}$ \\
$\omega$ & Argument of periastron (deg) & $159_{-56}^{+70}$ & $140_{-40}^{+20}$\\
$K$ & RV semiamplitude (m/s) & $8.6_{-1.8}^{+2.0}$ & $14.9_{-2.2}^{+2.1}$\\
$a/R_\star$ & Scaled semimajor axis & $10.1_{-0.3}^{+0.2}$ &$16.2_{-0.3}^{+0.3}$ \\
$i$ & Orbital inclination (deg) & $87.52_{-0.59}^{+1.14}$ & $89.28_{-0.37}^{+0.35}$ \\
$a$ & Semimajor axis (au) & $0.051_{-0.002}^{+0.002}$ & $0.061_{-0.001}^{+0.001}$ \\
$R_p$ & Planet radius ($R_{\oplus}$) & $6.7_{-0.2}^{+0.2}$ & $8.8_{-0.1}^{+0.1}$ \\
$M_p$ & Planet mass ($M_{\oplus}$) & $20.7_{-4.3}^{+4.8}$ & $37.6_{-5.5}^{+5.4}$ \\
$\rho_p$ & Planet mean density (g/cm$^{3}$) & $0.38_{-0.09}^{+0.10}$ & $0.31_{-0.05}^{+0.05}$ \\
\hline
$q_1^{\rm ESPRESSO}$ & ESPRESSO linear limb darkening parameter & $0.74_{-0.28}^{+0.18}$ & $0.54_{-0.32}^{+0.29}$\\
$q_2^{\rm ESPRESSO}$ & ESPRESSO quadratic limb darkening parameter & $0.65_{-0.37}^{+0.25}$ & $0.43_{-0.31}^{+0.36}$ \\
$\beta_{\rm ESPRESSO}$ & Intrinsic ESPRESSO stellar line width (km/s) & $4.4_{-2.0}^{+1.9}$ & $5.0_{-1.8}^{+1.8}$ \\
$\gamma_{\rm ESPRESSO}$ & ESPRESSO RV offset (m/s) & $4139_{-1}^{+1}$ & $-13004_{-1}^{+1}$\\
$\gamma_{\rm HARPS}$ & HARPS RV offset (m/s) & $4144_{-1}^{+1}$ & $-13011_{-2}^{+1}$ \\
$\gamma_{\rm PFS}$ & PFS RV offset (m/s) & $-1_{-2}^{+4}$ & - \\
$\gamma_{\rm FEROS}$ & FEROS RV offset (m/s) & $1_{-6}^{+5}$ & - \\
$\gamma_{\rm CORALIE1}$ & CORALIE1 RV offset (m/s) & - & $-13013_{-4}^{+4}$ \\
$\gamma_{\rm CORALIE2}$ & CORALIE2 RV offset (m/s) & - & $-12988_{-8}^{+8}$\\
$\sigma_{\rm ESPRESSO}$ & ESPRESSO RV jitter (m/s) & $0.1_{-0.1}^{+0.5}$ & $0.1_{-0.1}^{+0.6}$\\
$\sigma_{\rm HARPS}$ & HARPS RV jitter (m/s) & $0.6_{-0.6}^{+3.9}$ & $0.2_{-0.2}^{+1.3}$ \\
$\sigma_{\rm PFS}$ & PFS RV jitter (m/s) & $4.6_{-4.4}^{+5.1}$ & - \\
$\sigma_{\rm FEROS}$ & FEROS RV jitter (m/s) & $15.9_{-4.9}^{+6.1}$ & - \\
$\sigma_{\rm CORALIE1}$ & CORALIE1 RV jitter (m/s) & - & $0.4_{-0.4}^{+6.5}$ \\
$\sigma_{\rm CORALIE2}$ & CORALIE2 RV jitter (m/s) & - & $0.2_{-0.2}^{+3.5}$ \\
\hline
$q_1^{\rm TESS}$ & TESS linear limb darkening parameter & $0.36_{-0.18}^{+0.25}$ & $0.34_{-0.11}^{+0.14}$ \\
$q_2^{\rm TESS}$ & TESS quadratic limb darkening parameter & $0.16_{-0.12}^{+0.25}$ & $0.29_{-0.12}^{+0.17}$ \\
$q_1^{r'}$ & $r'$ linear limb darkening parameter & $0.19_{-0.11}^{+0.19}$ & - \\
$q_2^{r'}$ & $r'$ quadratic limb darkening parameter & $0.28_{-0.21}^{+0.37}$ & -\\
$q_1^{i'}$ & $i'$ linear limb darkening parameter & $0.16_{-0.07}^{+0.11}$ & -\\
$q_2^{i'}$ & $i'$ quadratic limb darkening parameter & $0.6_{-0.34}^{+0.27}$ & -\\
$q_1^{I+Z}$ & $I+z$ linear limb darkening parameter & - & $0.73_{-0.18}^{+0.16}$ \\
$q_2^{I+Z}$ & $I+z$ quadratic limb darkening parameter & - & $0.08_{-0.05}^{+0.09}$ \\
$q_1^{NGTS}$ & $NGTS$ linear limb darkening parameter & - & $0.52_{-0.13}^{+0.14}$ \\
$q_2^{NGTS}$ & $NGTS$ quadratic limb darkening parameter & - & $0.44_{-0.11}^{+0.15}$\\
$\sigma^{\rm TESS}_{\rm Y1}$ & TESS Year 1 photometric jitter (ppm) & $19_{-16}^{+89}$ & $14_{-11}^{+74}$ \\
$\sigma^{\rm TESS}_{\rm Y3}$ & TESS Year 3 photometric jitter (ppm) & $12_{-9}^{+66}$ & $11_{-9}^{+73}$ \\
$\sigma^{\rm TESS}_{\rm Y5}$ & TESS Year 5 photometric jitter (ppm) & $16_{-13}^{+79}$ & - \\
$\sigma^{r'}_{\rm MOANA/ES}$ & MOANA $r'$ photometric jitter (ppm) & $3005_{-135}^{+147}$ & -\\
$\sigma^{r'}_{\rm CHAT}$ & CHAT $r'$ photometric jitter (ppm) & $898_{-59}^{+65}$ & -\\
$\sigma^{i'}_{\rm LCO/SAAO}$ & LCO/SAAO $i'$ photometric jitter (ppm) & $51_{-46}^{+215}$ & -\\
$\sigma^{i'}_{\rm LCO/CTIO}$ & LCO/CTIO $i'$ photometric jitter (ppm) & $16_{-13}^{+110}$ & -\\
$\sigma^{I+z}_{\rm TRAPPIST}$ & TRAPPIST 2014 $I+z$ photometric jitter (ppm) & - & $111_{-105}^{+705}$ \\
$\sigma^{I+z}_{\rm TRAPPIST}$ & TRAPPIST 2015 $I+z$ photometric jitter (ppm) & - &  $2320_{-138}^{+142}$\\
$\sigma^{NGTS}_{\rm EulerCAM}$ & EulerCAM $NGTS$ photometric jitter (ppm) & - & $593_{-87}^{+86}$ \\
\enddata
\end{deluxetable*}

In Figure \ref{fig:lambda_vs_teff}, we compare our results for HATS-38 b and WASP-139 b to the sample of Neptunes with measured sky-projected obliquity from TEPCat\footnote{\url{https://www.astro.keele.ac.uk/jkt/tepcat/}} \citep{Southworth2011}, plotting the sky-projected obliquity versus stellar effective temperature. In this plot, we can see that Neptunes in compact multiplanet systems are all consistent with being well-aligned, except for the notable case of HD 3167 c \citep{Dalal2019,Bourrier2021}. In turn, isolated Neptunes can be either polar or well-aligned. This dichotomy has recently been discussed by \citet{radzom2024}.

A related property of these planets is their residual eccentricities (potential eccentricity for HATS-38 b), which poses a problem for tidal circularization models, as these would predict circularization timescales generally shorter than their ages \citep[see, e.g.,][]{Correia2020}. Similarly to \citet{EspinozaRetamal2023}, we estimate the circularization timescale for HATS-38 b and WASP-139 b following \citet{Goldreich66}. Assuming a modified quality factor \citep[e.g.,][]{Ogilvie07} $Q'=10^{6}$ and pseudo-synchronous rotation of the planet, we get circularization timescales of $\sim1.4$ and $\sim2.3$ Gyr, for HATS-38 b and WASP-139 b respectively, which are timescales shorter than the estimated ages of the systems of $\sim 10$ and $\sim 6$ Gyr, continuing with the observed trend for polar Neptunes.

Next, we discuss some of the properties of HATS-38 b and WASP-139 b as well as the overall distribution of stellar obliquities of the close-in Neptunes, some of the possible origins of these planets, and put some constraints on the possibility of having companions. 

\subsection{A possible preponderance of polar Neptunes?}

To better quantify the possible preponderance of polar Neptunes in the sample, using the two new measurements we have provided here, we used a Hierarchical Bayesian model \citep[HBM; e.g.,][]{Hogg2010} for the underlying obliquity distribution \citep[e.g.,][]{Morton2014,Munoz2018}. In particular, we used the framework\footnote{\url{https://github.com/jiayindong/obliquity}} presented by \citet{Dong23}, which models the underlying distribution of $\cos{\psi}$ across an exoplanet population using a mixture model of two Beta distributions \citep[e.g.,][]{Gelman14}. This hierarchical Bayesian modeling framework allows the user to derive the true 3D obliquity distribution from observed sky-projected obliquities and has the capacity to capture anything from an isotropic distribution to a strongly bimodal population. In this work, we derived the $\cos{\psi}$ distribution only from the $\lambda$ measurements and did not include information about the stellar inclination.

Both Beta distribution components are modeled using 3 parameters each: $w$, $\mu$, and $\kappa$. The parameter $w$ describes the weight of the component, while $\mu$ and $1/\kappa$ correspond to the mean and variance of each beta distribution component. The greater the value of $\kappa$, the smaller the variance, i.e., the distribution is more concentrated. The $\mu$ and $\kappa$ parameters can be related to the typical $\alpha$ and $\beta$ parameters of a beta distribution: $\alpha = \mu \kappa$ and $\beta=(1-\mu)\kappa$. 

\begin{figure*}[t]
    \centering
    \includegraphics[width=\textwidth]{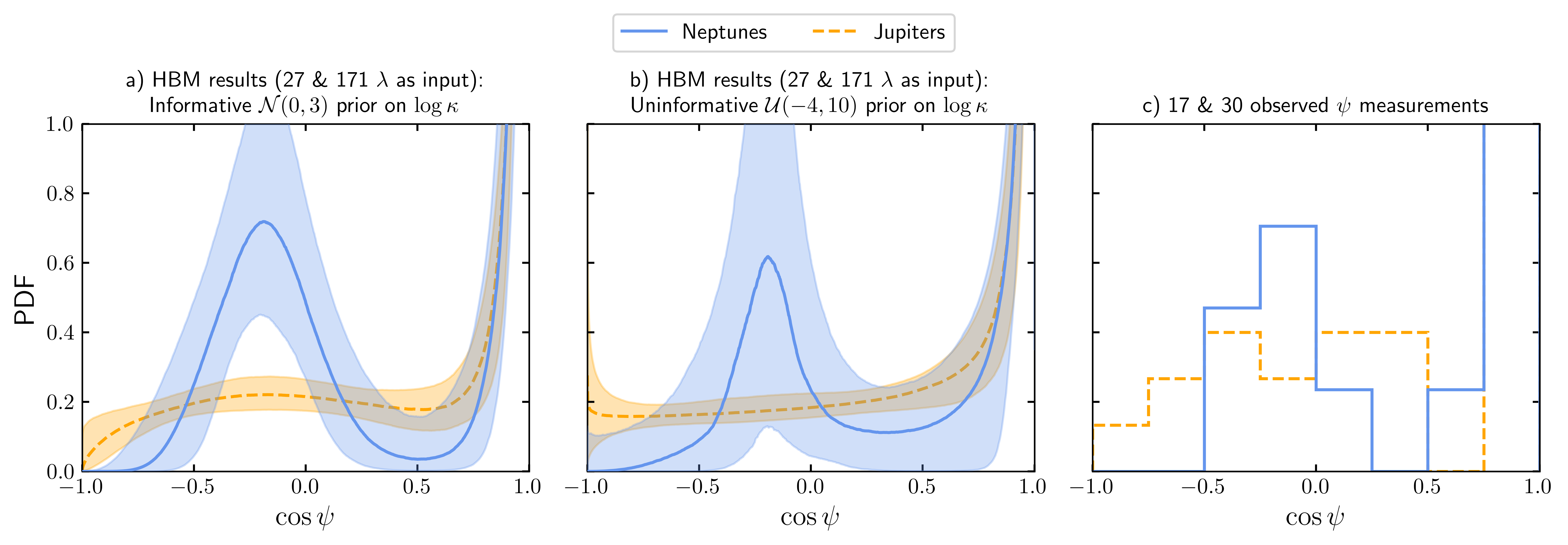}
    \caption{Stellar obliquity distribution of Neptunes (blue) and Jupiters (orange). a) Inferred distribution using the HBM with an informative $\mathcal{N}(0,3)$ prior on the $\log{\kappa}$ parameter. This inference was done following the methodology from \citet{Dong23} and based purely on observed sky-projected obliquities available in the literature---combined with the two sky-projected obliquity measurements presented here---without including information about the stellar inclination. The lines and shaded regions represent the median and $1\sigma$ uncertainties of the inferred distributions. b) Same as panel a but using a uninformative $\mathcal{U}(-4,10)$ prior on $\log{\kappa}$. c) Observed true 3D stellar obliquities from TEPCat for visual reference \citep[17 Neptunes and 30 Jupiters,][]{Southworth2011}. The true obliquity distribution also hints at a peak of polar Neptunes as inferred from the HBM modeling.}
    \label{fig:psi}
\end{figure*}

To derive the underlying $\cos{\psi}$ distribution, we ran two different hierarchical models with different priors on the compactness parameter $\kappa$: first, a model using informative priors on $\log \kappa$ following \citet{Dong23}, and second, a fit using uninformative priors on $\log \kappa$. For the first model, we followed \citet{Dong23}, and assumed an informative normal prior on $\log{\kappa}$ with a mean of 0 and a standard deviation of 3. Second, as an additional test---given the small size of the sample---we ran another model with an uninformative uniform prior between -4 and 10 for $\log{\kappa}$. In both cases, we used uniform priors on the location parameter $\mu$. The results of both of these model fits are shown in Table \ref{tab:beta}, and in Figure \ref{fig:psi}. For comparison, we also show in Figure \ref{fig:psi} the distribution of the true 3D obliquity measurements from TEPCat \citep[17 Neptunes and 30 Jupiters,][]{Southworth2011}.

We have found tentative evidence for a dichotomy in the orbital alignment of the sample of 27 Neptunes. Their orbits are preferentially well-aligned or polar. The relative weight of the two peaks depends on the priors we assumed, being $w_0=0.44_{-0.10}^{+0.11}$ and $w_1=0.56_{-0.11}^{+0.10}$ in the informative case, and $w_0=0.39_{-0.31}^{+0.24}$ and $w_1=0.61_{-0.24}^{+0.31}$ in the uninformative case. These weights suggest that about half of the Neptunes are expected to be in polar orbits while the other half is expected to be in well-aligned orbits. This is also reflected in the as-measured sky-projected obliquity sample: with the addition of our two new polar measurements, 13 Neptunes have $|\lambda|>50$ deg, and the remaining 14 have $|\lambda|<50$ deg. For both sets of priors, the polar peak is around $\cos{\psi} \sim -0.18$ while the aligned peak is around $\cos{\psi} \sim 1$. However, we see that the concentration of the polar peak and its uncertainty are strongly dependent on the priors used on $\log \kappa$, where the two different runs yield  $\log{\kappa_0}=3.08_{-1.47}^{+1.64}$ in the informative case and $\log{\kappa_0}=4.61_{-3.63}^{+5.15}$ in the uninformative case. For reference, $\log{\kappa_0}$ values of $\lesssim1.0$ correspond to isotropic distributions as seen from the $\log{\kappa_0}$ values of $1.24_{-0.92}^{+1.11}$ and $0.58_{-0.91}^{+3.26}$ for the Jupiter population in the informative and non-informative cases, respectively (see Table \ref{tab:beta}). If we define a $\log(\kappa_0) = 1$ to represent an isotropic distribution, we see that the Neptunes are $\sim$1.5$\sigma$ non-isotropic in the informative case, and consistent with isotropic at $\sim$1$\sigma$ in the non-informative case.

In contrast, the $\cos{\psi}$ distribution of Jupiters (171 systems, representing the entire sample with no cut in $T_{\rm eff}$) is not sensitive to the prior assumptions. Figure \ref{fig:psi} shows that the distribution has a peak at 0 deg representing the well-aligned systems, with an almost isotropic tail and no significative clustering at 90 deg. However, the respective weights of each beta component are different for the different priors. In the informative case, we have $w_0=0.33_{-0.06}^{+0.07}$ and $w_1=0.67_{-0.07}^{+0.06}$, suggesting that $\sim1/3$ of the hot Jupiters are in misaligned orbits and the other $\sim2/3$ are in well-aligned orbits. These results are consistent with the values derived by \citet{Dong23} for the sample of exoplanets from \citet{Albrecht2022}, which is dominated by hot Jupiters. However, with the uninformative priors, the weights are less constrained with $w_0=0.41\pm0.40$ and $w_1=0.59\pm0.40$ maybe suggesting that we could model the distribution using only 1 component.

The observed difference in the obliquity distribution for the sample of Neptunes and Jupiters---with the former showing a possible preference for a polar + aligned obliquity dichotomy, while the latter shows a preference for an aligned + isotropic dichotomy---likely suggests that hot Jupiters and hot Neptunes could have formed through different formation channels. However, the sample of Neptunes is small compared to the one of Jupiters (27 versus 171 systems), and the exact parameters of the bimodal distribution are different with different priors. According to \citet{Dong23}, in their tests, prior distribution choices did not strongly affect the derived distributions for samples of more than $\sim50$ systems. Therefore, more measurements of the stellar obliquity of Neptunes will be required to further confirm or rule out these results. 

Finally, we acknowledge that there are some biases that can affect these results given the small size of the Neptunes sample. Most of the Neptune obliquity measurements in our sample are measured via the RM effect (25 out of 27), except Kepler-30 b \citep[$\lambda=4\pm10$ deg,][]{Sanchis-Ojeda2012} and TOI-3884 b \citep[$\lambda=75\pm10$ deg,][]{Libby-Roberts2023} that were measured via starspost crossing events. As discussed in \citet{Siegel2023}, the RM effect is intrinsically biased towards either well-aligned or retrograde orbits and is biased against the detection of polar orbits due to their lower RM amplitudes. This bias \textit{against} polar orbits suggests that our result of detecting a candidate peak at polar orbits is robust---and, possibly, an underestimation. Future efforts like this one, and other ongoing programs like the Desert-Rim Exoplanets Atmosphere and Migration \citep[DREAM,][]{Bourrier2023}, will be crucial for increasing the size of the sample and understanding the formation and evolution of the hot Neptunes.

\subsection{Companions}\label{sec:companions}

The baseline of the RV measurements used in this work for HATS-38 and WASP-139 are $\sim7$ and $\sim15$ years respectively, and both residuals show no clear evidence of long-period companions (see panel c in Figures \ref{fig:HATS38} and \ref{fig:WASP139}). In order to estimate what kind of planetary companions can be ruled out by the current RVs, we performed population synthesis injection tests. For a synthetic population of possible companions, we estimated their RV signals and checked if they were consistent with the residuals of the RV model for HATS-38 b and WASP-139 b.

In both cases, we considered planetary companions with masses between 0.3 and 50 $M_J$. The eccentricity of the planets was drawn from a uniform distribution between 0.2 and 0.5, considering that cold Jupiters tend to be moderately eccentric \citep[see, e.g.,][]{Kipping13b}. The cosine of the orbital inclinations of the companions was drawn uniformly between 0 and 1\footnote{Similar results are obtained with the extreme case where all companions are edge-on.}. Finally, the semimajor axes were drawn uniformly from 0.5 and 100 au. With those distributions, we estimated the RV signals of different companions. Figure \ref{fig:companions} shows regions in the mass-semimajor axis plane ruled out at different confidence levels by the RV residuals. In both systems, the RVs rule out the existence of any planetary companions with masses between 0.3 and 50 $M_J$ within $\sim10$ au at $5\sigma$ confidence level.

Additionally to these possible planetary companions, we note that both stars appear to have stellar companions. \citet{Jordan2020} reported a possible M-dwarf companion to HATS-38, with a mass of 0.1 $M_\odot$ and a current projected separation of $\sim2100$ au. However, the source is not included in the Gaia catalog, given its low expected G magnitude of 23. So it can also be an extragalactic source, an earlier M-dwarf star that is in the background of HATS-38, or a foreground brown dwarf. As for WASP-139, \citet{El-Badry21} reported a comoving M-dwarf companion with a mass of 0.1 $M_\odot$ at a current projected separation of $\sim6400$ au. In both cases, the M-dwarf companion could have influenced the architecture of the system, as discussed in the next Section. 

\begin{deluxetable*}{lc|ccc|ccc|r}
\tablecaption{Summary of posteriors for the hyperparameters of the derived two-component $\cos{\psi}$ distribution. \label{tab:beta}}
\tablecolumns{9}
\tablewidth{0pt}
\tablehead{ & & \multicolumn{3}{c|}{Misaligned Component} & \multicolumn{3}{c|}{Well-aligned Component} & \\
Sample & Systems&$w_0$ & $\mu_0$ & $\log{\kappa_0}$ & $w_1$ & $\mu_1$ & $\log{\kappa_1}$& Reference}
\startdata
\textbf{Prior on $\log \kappa$ of $\mathcal{N}(0,3)$}$^a$ & &  &  &  & &  &  & \\
Neptunes & 27 & $0.44_{-0.10}^{+0.11}$ & $0.41\pm0.05$ & $3.08_{-1.47}^{+1.64}$ & $0.56_{-0.11}^{+0.10}$ & $0.97_{-0.03}^{+0.01}$ & $3.01_{-2.17}^{+1.78}$ & This work\\
Jupiters & 171& $0.33_{-0.06}^{+0.07}$ & $0.47_{-0.07}^{+0.08}$ & $1.24_{-0.92}^{+1.11}$ & $0.67_{-0.07}^{+0.06}$ & $0.98_{-0.01}^{+0.01}$ & $3.01_{-1.80}^{+1.67}$ & This work\\
A2022$^b$ & 161& $0.281\pm0.085$ & $0.434\pm0.088$ & $1.44\pm1.33$ & $0.719\pm0.085$ & $0.976\pm0.022$ & $2.65\pm1.88$ & D\&FM2023$^c$\\
\hline
\textbf{Prior on $\log \kappa$ of $\mathcal{U}(-4,10)$}$^a$ & &  &  &  & &  &  & \\
Neptunes & 27 & $0.39_{-0.31}^{+0.24}$ & $0.40_{-0.14}^{+0.12}$ & $4.61_{-3.63}^{+5.15}$ & $0.61_{-0.24}^{+0.31}$ & $0.94_{-0.23}^{+0.05}$ & $1.61_{-2.08}^{+5.61}$ & This work\\
Jupiters & 171 & $0.41\pm0.40$ & $0.56_{-0.38}^{+0.22}$ & $0.58_{-0.91}^{+3.26}$ & $0.59\pm0.40$ & $0.99_{-0.19}^{+0.01}$ & $3.88_{-3.99}^{+5.13}$ & This work\\
\enddata
\tablecomments{$^a$ $\mathcal{N}(\mu,\sigma)$ denotes a normal prior with median $\mu$ and standard deviation $\sigma$ while $\mathcal{U}(a,b)$ denotes a uniform prior between $a$ and $b$, $^b$\citet{Albrecht2022}, $^c$\citet{Dong23}.}
\end{deluxetable*}

\subsection{Possible Origin of the Polar Orbits}\label{sec:origin}

We discuss some of the proposed explanations for the origin of the polar orbits of hot Neptunes and whether these are able to explain not only the stellar obliquities of HATS-38 b and WASP-139 b but also their residual eccentricities (potential eccentricity in the case of HATS-38 b) and constraints on long-period companions. We argue that high-eccentricity migration from $\gtrsim 2$ au is a possible formation pathway for both systems, while primordial misalignments can be an option for HATS-38 b if its orbit is circular. In-situ excitation of obliquities and eccentricities is ruled out by our observational constraints.

\paragraph{Primordial disk misalignment}
Highly misaligned orbits can be produced through a primordial tilt of the protoplanetary disk by gravitational torques from a binary stellar companion \citep[e.g.,][]{Batygin2012,Lai2014,Zanazzi2018}. This scenario seems possible given that both stars appear to have distant M-dwarf companions at projected separations of $\sim2100$ au and $\sim6400$ au (Section \ref{sec:companions}). However, at these large separations a precession cycle of the disk is long\footnote{For reference, a protoplanetary disk extending out to 200 au would complete a precession cycle in $\sim 15$ Myr due to a M-dwarf companion with a semi-major axis of 2000 au (e.g., Equation (12) in \citealt{Zanazzi2018}).} compared to the disk dispersal timescale to resonantly excite the stellar obliquities \citep{Zanazzi2018}.  Alternatively, the stellar obliquity may be excited by the disk through magnetic disk torquing in a young accreting star \citep{Lai2011}. Although these may be plausible processes to misalign the parent disks of planets to the equator of their host stars, they are unable to explain the eccentric orbits. Given that we can not tell if the orbit of HATS-38 b is eccentric or not, these processes might be a likely explanation for its polar orbit only if the orbit is circular. If the orbit is eccentric, like in the case of WASP-139 b, these processes are less likely.

\begin{figure*}[t]
    \centering
    \includegraphics[width=\textwidth]{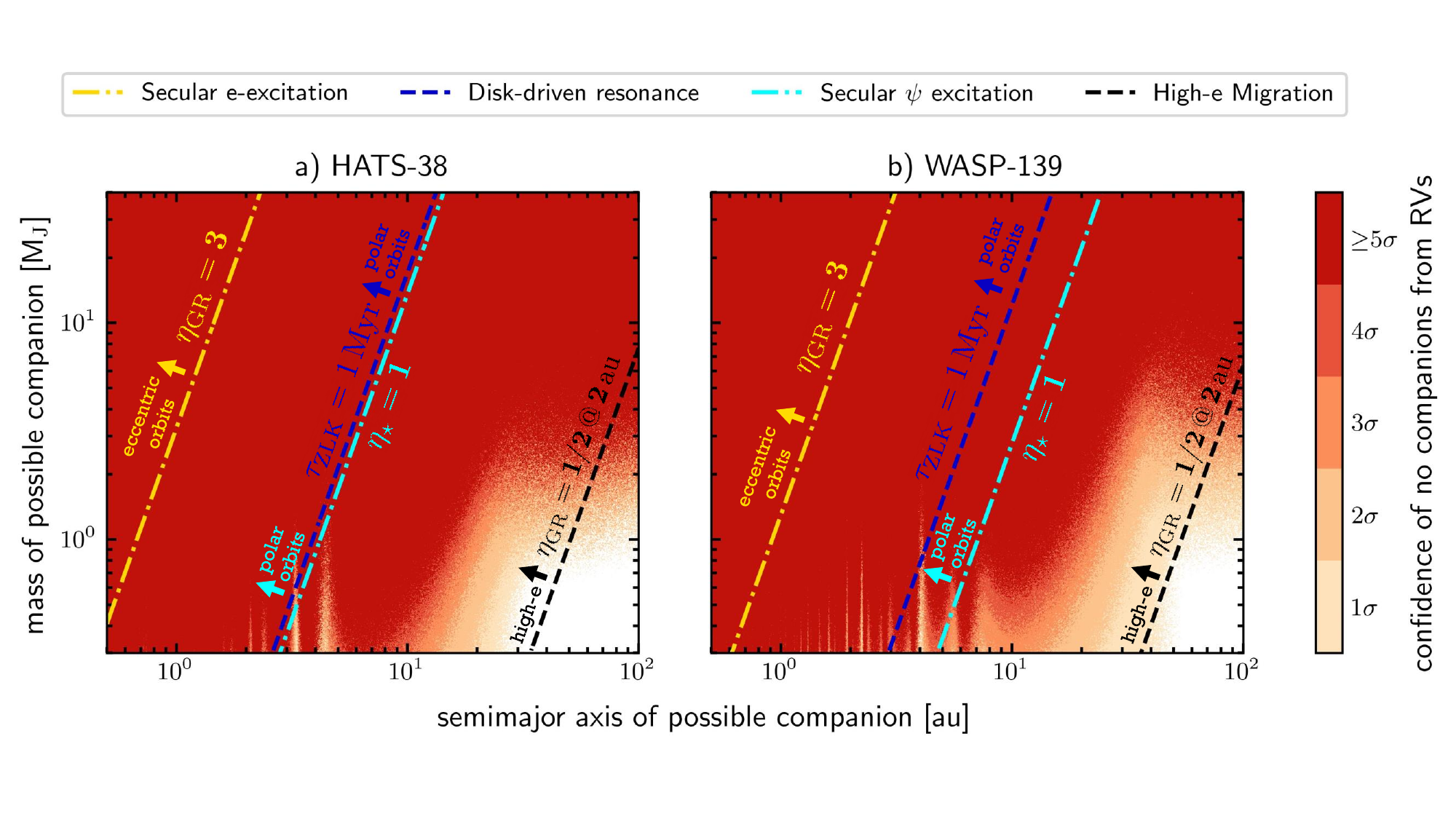}
    \caption{Mass versus semimajor axis diagram for the possible outer companions to a) HATS-38 b and b) WASP-139 b. Different colors indicate regions where we can rule out the companion at different $\sigma$ levels based on the RV residuals. \textbf{Yellow dotted-dashed lines} show $\eta_{\rm GR}=3$ (Equation \ref{eq:eta_gr}), demarking the regions where the companion can excite the eccentricities secularly. \textbf{Cyan dotted-dashed lines} show $\eta_\star=1$ (Equation \ref{eq:eta_star}) for $J_2=10^{-7}$ (corresponding to rotational periods of $\sim20-30$ days), demarking the regions where the companion can secularly excite obliquities. \textbf{Blue dashed lines} show $\tau_{\rm ZLK}= 1$ Myr (Equation \ref{eq:tau_zlk}), demarking the regions where the companion assisted by a dispersing protoplanetary disk can resonantly excite obliquities (adiabatic resonance capture). \textbf{Black dashed lines} show $\eta_{\rm GR}=1/2$, demarking the regions where the companion can induce high-eccentricity migration if it started at $a_{\rm b,0} = 2$ au. For any of the mechanisms discussed above to play a role, the companion needs to reside above the respective line; for companions below the line, the respective mechanism is excluded. From both figures, we see that the current RV observations are only compatible with the high-eccentricity migration scenario.}
    \label{fig:companions}
\end{figure*}

\paragraph{In-situ planet-planet scattering} The excited orbital state of planet b could have resulted from gravitational instabilities near its current orbital separation. As no other planets (with similar or large masses) are detected, such planets should have been ejected, leaving planet b in an eccentric/inclined orbit. We quantify the likelihood of this scenario by the Safronov-like parameter $\theta$ defined as the ratio of the escape velocity  of the planet  ($V_{\rm esc}$) and the orbital velocity ($V_{\rm orb}$):
\begin{equation}
\theta^2\equiv\frac{V_{\rm esc}^2}{V_{\rm orb}^2}=\frac{2M_p}{M_\star}\cdot\frac{a}{R_p}.
\end{equation}
For HATS-38 b and WASP-139 b, we obtain $\theta^2\sim0.02$ and $\sim0.04$, respectively, implying in both cases that scattering leads mainly to planetary collisions, not ejections. Thus, the excitation of eccentricities/inclinations becomes unlikely at the planet's orbital separation \citep{PTR_2014_scattering}.

\paragraph{In-situ excitation by a distant companion}
A distant giant planet companion (planet "c") can secularly tilt the planetary orbit and drive eccentricity growth on a time scale given by
\begin{equation}\label{eq:tau_zlk}
\tau_{\rm ZLK}=\frac{2P_{\rm b}^2}{3\pi P_{\rm c}}\frac{m_c}{M_\star}\left(\frac{a_c}{a_b}\right)^3,
\end{equation}
where ZLK stands for von Zeipel-Lidov-Kozai \citep{vonzeipel1910,lidov62, kozai62}.
The eccentricity growth can be quenched by relativistic apsidal precession (rate defined by $\dot{\omega}_{\rm GR}$) and its strength relative to the two-planet interaction can be quantified by the dimensionless parameter as \citep[e.g.,][]{FT2007,Liu2015}
\begin{equation}\label{eq:eta_gr}
    \eta_{\rm GR} =\dot{\omega}_{\rm GR}\tau_{\rm ZLK}= \frac{4GM_\star}{c^2}\frac{a_c^3(1-e_c^2)^{3/2}}{a_b^4}\frac{M_\star}{m_c}.
\end{equation}
For $\eta_{\rm GR}>3$ eccentricity excitation is quenched. In Figure \ref{fig:companions}, we show that the allowed regions for HATS-38 b and WASP-139 b (yellow dot-dashed lines) are largely excluded by long-term RV measurements.

The inclinations can, in turn, be secularly excited by an inclined companion and compete with the quadrupolar gravitational field with the host star parametrized by $J_2$.
We define the relative stellar quadrupole with respect to the two-planet interaction as in \citet{Petrovich2020}
\begin{equation}\label{eq:eta_star}
    \eta_{\star} \equiv\frac{2J_2M_\star}{m_c}\frac{R_\star^2a_c^3(1-e_c^2)^{3/2}}{a_b^5}.
\end{equation}
For $\eta_\star>1$ the distant planet c is unable to drive the growth of the stellar obliquity. In both panels of Figure \ref{fig:companions} we plot $\eta_\star = 1$ assuming $J_2=10^{-7}$ and the observed stellar masses and radii, and find that we can rule out the presence of a planetary companion in most of the regions where the necessary condition is satisfied. 

A final possibility we assess is that the secular excitation occurred early on, assisted by the dispersal of the birth protoplanetary disk, as proposed by \citet{Petrovich2020}. Here, the disk promotes the capture into an inclination secular resonance (with possible eccentricity excitation) if the dispersal of the disk is longer than the adiabatic timescale, which is defined as $\tau_{\rm adia}=\tau_{\rm ZLK}(1+\eta_\star)^{1/3}/I_{c,0}^{4/3}$ where $I_{c,0}$ is the initial inclination of planet c relative to the disk. Since $I_{c,0}\ll1$, we can comfortably assume that $\tau_{\rm adia}\gg \tau_{\rm ZLK}$ and use this bound to compare it with a typical dispersal timescale of the disk of $1$ Myr. In Figure \ref{fig:companions} we show the $\tau_{\rm ZLK}=1$ Myr, which again, the condition $\tau_{\rm ZLK}<1$ Myr is largely ruled out by the radial velocity constraints.

In summary, the secular excitation of the planet's stellar obliquity at its current orbital separation is largely disfavored by our RV constraints.

\paragraph{High-eccentricity migration driven by a distant companion}
We have ruled out various mechanisms that can explain the excited orbital states of the Neptunes at their current orbital distances, except for a primordial misalignment in the case of a circular orbit for HATS-38 b. Finally, we explore the possibility of high-eccentricity tidal migration: the excitation of the stellar obliquities and eccentricities at wider orbital separations, followed by inward tidal migration that circularizes the orbits over time.

High-eccentricity tidal migration driven by planet-planet scattering accompanied by secular interactions \citep[e.g.,][]{Nagasawa2008} has been proposed as an explanation for the eccentric and misaligned orbits of hot Neptunes \citep[see, e.g.,][]{Bourrier2018,Correia2020,Stefansson22}, including recent applications to the systems HAT-P-11 \citep{Lu_ZLK_2024} and WASP-107 \citep{YU2024}. But, where should the migration have started to still be consistent with the excluded regions by our RV measurements?

A planet b can acquire an extreme eccentricity to allow for migration only if $\eta_{\rm GR}\lesssim1$. For reference, a maximum eccentricity of $e_{\rm max}\gtrsim0.95$ can be achieved only if $\eta_{\rm GR}>1/2$. In Figure \ref{fig:companions} we plot $\eta_{\rm GR} = 1/2$ for an initial semimajor axis of the transiting planet of $a_{\rm b,0} = 2$ au, which coincides with regions excluded at only $\gtrsim  1\sigma$ for companions with masses $\lesssim 3$ $M_J$.  Given the steep dependence with the initial semi-major axis ($\eta_{\rm GR}\propto a_{\rm b,0}^{-4}$), we can conclude that ZLK migration could still be consistent with excluded regions if $a_{\rm b,0}\gtrsim2$ au. 

If no companions are found, we are left with the wide-orbit binary companions as the driver of ZLK migration (e.g., \citealt{FT2007,naoz2012,petrovich2015}). Though ZLK migration starting in itself does not predict an excess of polar planets, the combination of disk tilting due to the binary and subsequent ZLK migration can produce an excess of polar and retrograde hot planets \citep{Vick2023}.
In this scenario, migration should have started from an even wider orbit to not be quenched by GR ($\eta_{\rm GR}<1/2$ for $a_{\rm b,0}\gtrsim9$ au with $m_c=0.1$ $M_\odot$ and $a_c=2000$ au, see Section \ref{sec:companions}). 
Additionally, the migrating planets should lack any other planetary companions to avoid quenching the ZLK cycles, a state that could be achieved by planet-planet scattering.

\section{Conclusions and Summary}\label{sec:conclusion}

In this work, we have presented new observations of the hot Neptunes HATS-38 b and WASP-139 b. We have jointly analyzed ESPRESSO observations of the RM effect produced by the transiting planets, with photometry and out-of-transit radial velocities using the publicly available code \texttt{ironman}. We have concluded that:

\begin{itemize}
    \item HATS-38 b has a polar and potentially eccentric orbit. We have measured a sky-projected obliquity $\lambda = -108_{-16}^{+11}$ deg for the orbit, and an eccentricity of $e= 0.112_{-0.070}^{+0.072}$ that is compatible with 0 given our data. Future RV monitoring and/or eclipse observations will be needed to robustly distinguish between an eccentric and circular orbit for this planet.

    \item WASP-139 b has a polar and eccentric orbit. We have measured a sky-projected obliquity $\lambda = -85.6_{-4.2}^{+7.7}$ deg, and an eccentricity $e=0.103_{-0.041}^{+0.050}$.
    
    \item Neither systems show clear evidence of long-period companions in long-term RV observations. Based on the residual RVs we have ruled out the presence of planetary companion with masses $\sim 0.3-50$ $M_J$ within $\sim10$ au at $5\sigma$ confidence level. 

    \item Both planets join a growing population of isolated low-density Neptunes in polar orbits, which includes HAT-P-11 b \citep{Sanchis-Ojeda2011}, GJ 436 b \citep{Bourrier2018,Bourrier2022}, WASP-107 b \citep{Dai2017,Rubenzahl2021}, and GJ 3470 b \citep{Stefansson22}.
    
    \item A possible explanation for the polar and eccentric (potentially eccentric in the case of HATS-38 b) orbits of these planets is high-eccentricity tidal migration produced by secular interactions from yet undetected outer planetary companions or the observed wide-orbit stars. If HATS-38b has a circular orbit, the polar orbit could also be explained with formation in a primordially misaligned disk. In contrast, we disfavor in-situ excitation of inclinations and eccentricities for either system, as this would require companions at orbital distances we have ruled out with out-of-transit RVs.

    \item Through hierarchical Bayesian modeling of sky-projected obliquity measurements, we found suggestive evidence for a preponderance of polar orbits of hot Neptunes. In contrast, and similar to recent work by \citet{Siegel2023} and \citet{Dong23}, we do not observe significant clustering of hot Jupiters in polar orbits. However, we note that the exact obliquity distribution of Neptunes is sensitive to the choice of priors, highlighting the need for additional obliquity measurements of Neptunes to robustly compare the hot Neptune obliquity distribution to hot Jupiters.
\end{itemize}

\section*{Acknowledgments}
We would like to thank Coel Hellier for the TRAPPIST and EulerCAM light curves of WASP-139. We thank Jiayin Dong for her help with the implementation of the Bayesian framework to derive the true 3D obliquity distributions. We also thank the anonymous referee for their thoughtful comments and suggestions that strengthened the manuscript.
J.I.E.R and C.P. acknowledge support from ANID BASAL project FB210003.
J.I.E.R. acknowledges support from ANID Doctorado Nacional grant 2021-21212378.
C.P. acknowledges support from FONDECYT Project 1210425, CASSACA grant CCJRF2105, and ANID+REC Convocatoria Nacional subvenci\'on a la instalaci\'on en la Academia convocatoria 2020 PAI77200076.
A.J.\ and R.B.\ acknowledge support from ANID -- Millennium  Science  Initiative -- ICN12\_009 and AIM23-0001.
R.B. acknowledges support from FONDECYT Project 1241963.
A.J.\ acknowledges support from FONDECYT project 1210718.
J.P.L. acknowledges co-funding by the European Union (ERC, FIERCE, 101052347), by FCT through national funds, and by FEDER through COMPETE2020 - Programa Operacional Competitividade e Internacionalização grants UIDB/04434/2020 and UIDP/04434/2020.
M.T.P. acknowledges support from the ANID Post-doctoral fellowship 3210253.

\bibliography{sample631}{}
\bibliographystyle{aasjournal}

\end{document}